\documentclass[10pt,a4paper]{article}%
\usepackage{epsfig}
\usepackage{amsmath}
\usepackage{amsfonts}
\usepackage{amssymb}
\usepackage{graphicx}

\begin{document}

\begin{flushright}
{\small LPTh-Ji 10/006}
\end{flushright}

\begin{center}
\vspace{2cm}

{\Large \textbf{\textit{Symmetry Restoration at High Temperature in
Little Higgs Models?}}}

\vspace{1.3cm}

\textbf{Amine Ahriche}

\vspace{0.6cm}

\textit{Laboratory of Theoretical Physics, Department of Physics,
University of Jijel, PB 98 Ouled Aissa, DZ-18000 Jijel, Algeria.}

\textit{Email: ahriche@univ-jijel.dz}
\end{center}

\vspace{1.2cm}

\hrule \vspace{0.5cm} {\normalsize \textbf{{\large {Abstract}}}}

\vspace{0.5cm}

In this work, we show that the apparent symmetry non-restoration at
high temperature for Little Higgs (LH), is not intrinsic feature for
LH. We show that when including such dominant thermal corrections,
the EW symmetry gets restored.

\vspace{0.8cm} \textbf{Keywords}: Little Higgs, symmetry breaking,
symmetry restoration at high temperature. \vspace{0.5cm}

 \vspace{0.5cm} \hrule\vspace{2cm}

\section{Introduction}

Recently the little Higgs (LH) mechanism has been proposed as a way to
stabilize the weak scale from the radiative corrections of the Standard
Model. In little Higgs models, the Standard Model Higgs doublet appears as
pseudo-Goldstone in a global symmetry breaking; and it is kept light by
approximate non-linear symmetries \cite{LH}. The little Higgs mechanism
requires that two separate couplings communicate to the Higgs sufficient
breaking of the non-linear symmetry to generate a Higgs mass. The weak scale
is radiatively generated two loop factors below a cut-off estimated using
naive dimensional analysis (\textit{NDA}), $\Lambda\sim4\pi f$, where $%
f\sim1-10$ TeV, is the vev that breaks the global symmetry. The Higgs mass,
in this setup, is protected from quadratic corrections at one-loop above the
cut off. The scalar fields correspond to the broken generators of the global
symmetry; and the scalar potential is just a combination of effective
operators whose gauge and Yukawa origins. The electroweak symmetry breaking
(EWSB) is triggered by large one-loop Yukawa contributions \cite{LH,LH2}.
There are many variants of this scenario depending on the chosen global
symmetry. The famous version is the so-called Littlest Higgs \cite{LH},
which is based on a global $SU(5)$ symmetry spontaneously broken to $SO(5)$.
These models are phenomenologically consistent \cite{phlh}.

However the EW symmetry restoration at temperature seems to be problematic
as shown for the Littlest Higgs \cite{ELR}. This can be understood due to
the fact that thermal corrections ($\sim T^{2}/12$); and quadratic
corrections ($\sim3\Lambda^{2}/16\pi^{2}$) are generated at one-loop from
the same interactions in any gauge theory. This means that the Higgs thermal
corrections will cancel each other also; and the electroweak symmetry will
not be restored at high temperature. Indeed, it was shown in Ref. \cite{ELR}%
, that above such critical temperature $T_{c}\sim f$, the thermal
corrections become negative and the absolute minimum ($\left\langle
h\right\rangle \neq0$), gets deeper at higher temperatures instead of being
relaxed to zero.

In this work, taking the Littlest Higgs as an example, we will show that
this unusual behavior of the effective scalar potential is realistic but due
to incomplete computations, and the symmetry can be restored at high
temperatures by taking some ignored corrections. In the next section, we
briefly introduce the Littlest Higgs model. In the third section, we discuss
the electroweak symmetry breaking and the non-restoration at high
temperature. In section 4, we introduce some important higher order
corrections to the effective potential and show that the symmetry is
restored at high temperatures. Finally we give our conclusion.

\section{Basics of the Littlest Higgs}

The Littlest Higgs model \cite{LH} is based on an $SU(5)/SO(5)$ nonlinear
sigma model, where the $SU(5)$ symmetry is spontaneously broken down to $%
SO(5)$ by the vacuum expectation value of a $5\times5$ symmetric matrix
scalar field%
\begin{equation}
\Sigma_{0}=\left(
\begin{array}{ccc}
0 & 0 & \mathbf{1}_{2\times2} \\
0 & 1 & 0 \\
\mathbf{1}_{2\times2} & 0 & 0%
\end{array}
\right) .
\end{equation}
The breaking of the global $SU(5)$ symmetry results 14 Goldstone bosons: 4
are identified as the Standard Model (SM) Higgs doublet, 6 as complex
triplet and 4 as the Goldstone bosons that give masses to the new heavy
gauge bosons. These Goldstone bosons can be parameterized through the
nonlinear%
\begin{align}
\Sigma & =e^{i\Pi/f}\Sigma_{0}e^{i\Pi^{T}/f}=e^{2i\Pi/f}\Sigma_{0},
\label{Sig} \\
\Pi & =\pi^{a}X^{a}.  \nonumber
\end{align}
Here $X^{a}$ are the $SU(5)$ broken generators. The $SU(5)$ subgroup ($%
SU(2)\times U(1)$)$_{1}\times$($SU(2)\times U(1)$)$_{2}$; is gauged, its
diagonal subgroup being the SM electroweak group $SU(2)_{L}\times U(1)_{Y}$.
The gauge interactions can be seen from the kinetic term in the Lagrangian%
\begin{align}
{\mathcal{L}}_{\Sigma} & =\frac{f^{2}}{8}\mathrm{Tr}\left\{ \left( D_{\mu
}\Sigma\right) \left( D^{\mu}\Sigma\right) ^{+}\right\} ,  \label{k} \\
D_{\mu}\Sigma & =\partial_{\mu}\Sigma-i\sum_{i}\left[ g_{i}W_{i}^{a}\left(
Q_{i}^{a}\Sigma+\Sigma Q_{i}^{aT}\right) +g_{i}^{\prime}B_{i}\left(
Y_{i}\Sigma+\Sigma Y_{i}^{T}\right) \right] ,
\end{align}
and%
\begin{align}
Q_{1}^{a} & =\left(
\begin{array}{cc}
\sigma^{a}/2 &  \\
&
\end{array}
\right) ,~Q_{2}^{a}=\left(
\begin{array}{cc}
&  \\
& -\sigma^{a\ast}/2%
\end{array}
\right) ,  \nonumber \\
Y_{1} & =\mathrm{diag}\left( -3,-3,2,2,2\right) /10,~~Y_{2}=\mathrm{diag}%
\left( -2,-2,-2,3,3\right) /10;
\end{align}
are the gauge generators, $\sigma^{a}$ are the Pauli matrices. The gauge
couplings $g_{1,2}$\ and $g_{1,2}^{\prime}$\ are related to the SM ones as $%
1/g^{2}=1/g_{1}^{2}+1/g_{2}^{2}$ and $1/g^{\prime2}=1/g_{1}^{%
\prime2}+1/g_{2}^{\prime2}$. The quark sector involves a new heavy singlet
quark $U$, where the Yukawa interaction is given by
\begin{equation}
{\mathcal{L}}_{Y}=-\tfrac{\lambda_{1}}{2}f\bar{u}_{R}\mathrm{Tr}\left\{
\left( \Sigma_{1}\epsilon\Sigma_{1}^{T}\right) \left[ \chi\right]
^{T}\right\} -\lambda_{2}f\bar{U}_{R}U_{L}+h.c,
\end{equation}
with $\Sigma_{1}=\left\{ \Sigma_{im}\right\} $\ and $m,n=4,5$, $i,j=1,2,3$;\
$\left[ \chi\right] _{ij}=\epsilon_{ijk}\chi_{Lk}$ and $\chi_{L}^{T}=\left(
u_{L},b_{L},U_{L}\right) $, where $u$\ and $U$\ are mixtures of the top
quark $t$ and the new heavy quark $T$. Here $\lambda _{1,2}$ are related to
the top Yukawa coupling as $1/\lambda_{t}^{2}=1/\lambda_{1}^{2}+1/%
\lambda_{2}^{2}$.

In general, the neutral components of both doublet and triplet can develop a
real vev. Then the Coleman-Weinberg potential \cite{CW}, $V_{CW}(\Sigma)$,
is a result of such effective operators coming from two sources: gauge and
fermions interactions%
\begin{align}
V_{CW}(\Sigma) & =a_{V}f^{4}\left[ g_{i}^{2}\mathrm{Tr}\left\{ \left(
Q_{i}^{a}\Sigma\right) \left( Q_{i}^{a}\Sigma\right) ^{\ast}\right\}
+g_{i}^{\prime2}\mathrm{Tr}\left\{ \left( Y_{i}\Sigma\right) \left(
Y_{i}\Sigma\right) ^{\ast}\right\} \right]  \nonumber \\
& +\frac{a_{F}}{4}\lambda_{1}^{2}f^{4}\mathrm{Tr}\left\{ \left[
\Sigma
_{1}\epsilon\Sigma_{1}^{T}\right] \left[ \Sigma_{1}\epsilon\Sigma_{1}^{T}%
\right] ^{\dag}\right\} ,
\end{align}
where $a_{V}$, $a_{F}\sim\mathcal{O}\left( 1\right) $\ are unknown
parameters associated with the effective operators; their values
depend on the UV completion of the theory. The Coleman-Weinberg
potential is given in the $h$ and $\phi$\ directions by
\begin{equation}
V_{CW}\left( h\right) =\Theta f^{4}\sin^{4}\left( h/\sqrt{2}f\right)
,~V_{CW}\left( \phi\right) =\Theta f^{4}\sin^{2}\left( \sqrt{2}\phi/f\right)
,   \label{Vv}
\end{equation}
respectively, with $\Theta=a_{V}\left[
g_{1}^{2}+g_{2}^{2}+g_{1}^{\prime 2}+g_{2}^{\prime2}\right]
/4+a_{F}\lambda_{1}^{2}/2$. It is clear that since
$\Theta>0$, the theory ground state is stable in both of $h$- and $\phi $%
-directions; and the EW symmetry is not broken. But when including the
one-loop corrections (especially the Yukawa's), the EW symmetry gets broken
and the SM fields develop masses \cite{LH,LH2}.

\section{Symmetry Breaking and Restoration at High Temperature}

In Ref. \cite{ELR}, it was shown in figures (1) and (2); how the EW symmetry
gets broken in a slight way after including the one-corrections. While in
figure (3), it was shown how thermal corrections help to relax the minimum
to zero especially for low temperature values ($T\leq0.44f$); but the
maximum of the potential (at $h=\pi f/\sqrt{2}$) is getting down when
increasing the temperature until becomes degenerate together with the
absolute minimum at a critical temperature $T_{c}\sim0.96f$. Above this
temperature value ($T>f$), this new minimum gets deeper and deeper when
increasing temperature. This unusual behavior had lead to the conclusion
that symmetry can not be restored at high temperatures in these models.

The scalar potential behavior at high temperature that is mentioned in Ref.
\cite{ELR}\ can be understood by estimating the field-dependant masses,
especially the Yukawa's, within one period $h\in\lbrack0,\sqrt{2}\pi f]$.
Naively, the gauge field masses vary below the squared of the gauge
couplings in units of $f^{2}$, scalar masses vary below a combination of $%
a_{V}g_{i}^{2}$ and $a_{F}\lambda^{2}$ in units of $f^{2}$. These masses are
significantly small when compared with the two Yukawa eigenmasses which vary
between \{0$\sim\lambda_{t}^{2}$\} and \{$\lambda_{2}^{2}\sim%
\lambda_{1}^{2}+\lambda_{2}^{2}$\}. According to these values, the thermal
integral \cite{Th}, that involves $\left( m^{2}/T^{2}\right) $ as a
variable, will not be suppressed for all fields; and the unsuppressed
fermionic contributions, that have a large negative multiplicity ($%
n_{t}=n_{T}=-12$), will dominate the thermal scalar potential since they are
$T^{4}$-proportional. In other gauge theories, SM as an e.g.,\ most of the
fields contributions are suppressed for large values of the scalar field $h$%
, but this not the case here because the scalar sector is periodic.
Therefore one concludes that this behavior is a consequence of periodicity
of the non-linear nature of the scalar representation; in addition to the
largeness of negative fermionic contributions to the thermal effective
potential, that can not be compensated by other bosonic contributions. But
were all the significant contributions taken into account to obtain this
behavior?

\section{Thermal Corrections and Symmetry Restoration at High Temperature}

\begin{figure}[h]
\centerline{\psfig{file=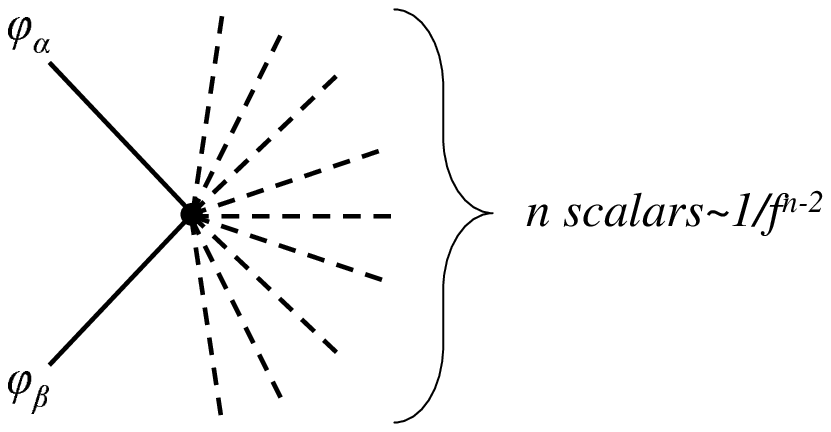,width=4.5cm,height=2.5cm}} \vspace{8pt}
\caption{\textit{The fields $\protect\varphi_{\protect\alpha}$ and $\protect%
\varphi_{\protect\beta}$ could be scalars, gauge fields or a
fermion-antifermion pair.}}
\label{nl}
\end{figure}

The nonlinear nature of the scalar sector implies the existence of a new
type of interactions like in Fig-\ref{nl}, when expanding the field matrix $%
\Sigma$ in powers of $1/f$ in the Lagrangian. \cite{next}

\begin{figure}[h]
\centerline{\psfig{file=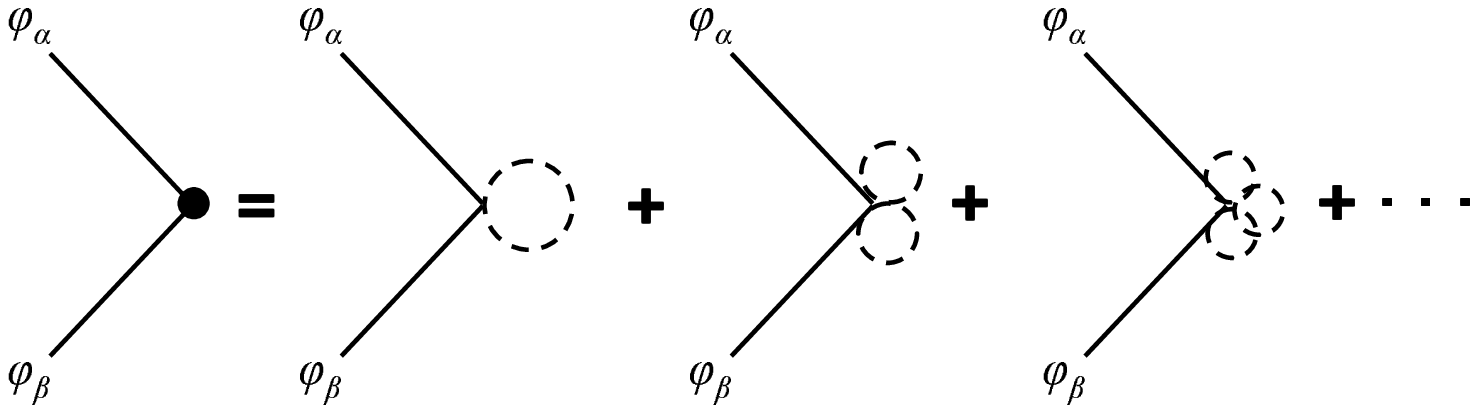,width=7.5cm,height=1.8cm}} \vspace{8pt}
\caption{\textit{Higher order loops corrections to the mass-squared matrix
element $M_{\protect\alpha\protect\beta}^{2}$, that come from the
interactions in Fig-\protect\ref{nl}.}}
\label{v2}
\end{figure}

\begin{figure}[h]
\centerline{\psfig{file=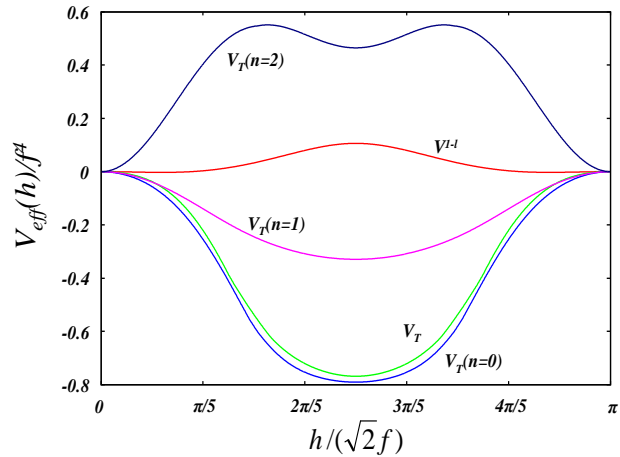,width=8cm,height=6cm}} \vspace{8pt}
\caption{\textit{The effective potential at T=0 (solid line), and at T=1,7f
computed in the standard way, and using the resumed thermal masses (\protect
\ref{mto}) taking into account 1-loop (n=0), 2-loop (n=1), and 3-loop (n=2)
corrections.}}
\label{Vt}
\end{figure}

These vertices which are proportional to $1/f^{n-2}$; could result higher
order loop corrections as shown in Fig-\ref{v2}. The corrections in Fig-\ref%
{v2}\ are not the only possible contractions of the vertices in Fig-\ref{nl}%
, but they are the dominant contributions since each scalar loop gives $%
T^{2}/12$. Therefore, they lead to a thermal correction to the field
mass-squared matrix element [$\alpha,\beta$], of the form%
\begin{equation}
m^{2}(T)\sim m^{2}+T^{2}\sum\nolimits_{n}c_{n}\left( T^{2}/f^{2}\right)
^{n},   \label{mto}
\end{equation}
where the zeroth order corresponds to the usual thermal corrections. The
computation of the\ $c_{n}$ parameters for each mass-squared matrix element
tends to determine the vertices of the interactions with scalar degrees of
freedom \cite{next}. This type of contributions (\ref{mto}); could be very
important at temperatures $T\gtrsim f$ . One can take these novel
corrections into account by doing such a resummation where the
field-dependant masses in the thermal integral \cite{Th}, could be replaced
by thermally corrected masses (\ref{mto}). In Fig-\ref{Vt}, we show how does
the inclusion of these novel corrections affect the thermal effective
potential behavior at $T=1.7f$ as an example.

As it is clear, it is enough to consider only the order ($n=2$) in (\ref{mto}%
) to see that the EW symmetry is restored at this temperature. Here is this
example, we have chosen the cut-off scale to be $\Lambda=1.3f$, if it is
taken to be the \textit{NDA} value $\Lambda=4\pi f$, then we need more
corrections ($n>2$) to show that the symmetry is restored, and for $n=2$,
the EW symmetry is restored at $T\simeq2.854f.$ This mean that the thermal
effective potential behavior shown in Ref. \cite{ELR}, is not intrinsic but
due the incomplete theory above such a validity scale.

\section{Conclusion}

In this work, we have shown that the symmetry non-restoration at high
temperature behavior for Little Higgs is not an intrinsic feature for LH,
but just due incomplete computation setup. By taking some high order
dominant corrections, we found that the symmetry is restored at high
temperature for LH.

\subsection*{Acknowledgements}

I want to thank the organizers for inviting me to this workshop.
This work was supported by the Algerian Ministry of Higher Education
and Scientific Research under the cnepru-project number
D01720090023.

\end{document}